\documentclass[
    twocolumn,
    showpacs,
    preprintnumbers,
    amsmath,
    amssymb,
    prc,
    superscriptaddress,
    floatfix
]{revtex4-2}
\usepackage[utf8]{inputenc}
\usepackage{graphicx}
\usepackage{hyperref}
\usepackage[english]{babel}

\newcommand{\ILNP}{Laboratory of Nuclear Problems, Joint Institute for Nuclear Research, RU-141980 Dubna, Russia}
\newcommand{\IAlm}{Institute of Nuclear Physics, KZ-050032 Almaty, Kazakhstan}
\newcommand{\IAst}{L.N.~Gumilyov Eurasian National University, KZ-010000 Nur-Sultan, Kazakhstan}
\newcommand{\IMSU}{Department of Physics, Moscow State University, RU-119991 Moscow, Russia}
\newcommand{\IDub}{Dubna State University, RU-141980 Dubna, Russia}

\begin{document}
\title{Resonant behavior of the \texorpdfstring{$pp \to \{p p\}_s \pi^0$}{pp -> \{pp\}s pi0} reaction at the energy \texorpdfstring{$\sqrt{s} = 2.65$}{sqrt(s) = 2.65}~GeV}
\author{D.~Tsirkov}\affiliation{\ILNP}
\author{B.~Baimurzinova}\email{baimurzinova@jinr.ru}\affiliation{\ILNP}\affiliation{\IAlm}\affiliation{\IAst}
\author{V.I.~Komarov}\affiliation{\ILNP}
\author{A.V.~Kulikov}\affiliation{\ILNP}
\author{A.~Kunsafina}\affiliation{\ILNP}\affiliation{\IAlm}\affiliation{\IAst}
\author{V.S.~Kurbatov}\affiliation{\ILNP}
\author{Zh.~Kurmanalyiev}\affiliation{\ILNP}\affiliation{\IAlm}\affiliation{\IAst}
\author{Yu.N.~Uzikov}\affiliation{\ILNP}\affiliation{\IMSU}\affiliation{\IDub}

\date{\today}

\begin{abstract}
We report on measurements of the differential cross section $d\sigma/d\Omega$ in the energy region of $\sqrt{s} = 2.5$--$2.9$~GeV of the reaction $pp \to \{pp\}_s \pi^0$ where $\{pp\}_s$ is a $^1\!S_0$ proton pair.
The experiment has been performed with the ANKE spectrometer at COSY-J\"ulich.
The data reveal a peak in the energy dependence of the forward $\{pp\}_s$ differential cross section at the energy $\sqrt{s} = 2.65$~GeV with the width $\Gamma = 0.26$~GeV, and also the cross section slope changes its sign in the region of the observed peak.
This is an indication for the excitation of a new $D(2650)$ resonant state, that may be a dibaryon resonance system consisting of two excited baryons $\Delta(1232)$ and $N^*(1440)$.
\end{abstract}

\maketitle

\section{Introduction}

One of the striking phenomena of modern particle physics is dibaryon resonances.
In the broad sense of this term, the dibaryon resonance means an enhancement of interaction at a specific energy in a dibaryon system, i.e., a hadron system with the baryon number $B = 2$.

Currently, there is no unambiguous and generally accepted understanding of the physical nature of the dibaryon resonance phenomenon, and its interpretation is developing in two main directions.
One is the meson-baryon approach (developed, e.g., in~\cite{Gal}), which assumes that dibaryon resonances represent the resonant amplification of the interaction between a nucleon and an excited baryon, or between two excited baryons.
Meson fields are the carrier of interaction in this case.
The other approach assumes that the effective degrees of freedom in the dibaryon resonances are quark degrees of freedom, and the excited dibaryon system is a quasi-bound six-quark state.
The symbiosis of these approaches is the idea of the dibaryon resonances as an ensemble of excited hadronic and quark components (for details see~\cite{Dong} and refs.\ therein).

Regardless of the microscopic nature of dibaryon resonances, they can be classified by analysing purely symmetric properties of hadron systems in SU(6) theory.
This classification was proposed by Dyson and Nguyen Xuong in~\cite{Dyson}.
The dibaryon states $D_{IJ}$ are defined in this classification by the isotopic spin $I$ and the total momentum $J$.

The authors~\cite{Dyson} had only three states for analysis: the deuteron $D_{01}$; the isotriplet state $D_{10}$ of an $np$ pair (or an $S$-wave $pp$ pair ${pp}_s$) and the $S$-wave $N\Delta$ resonance $D_{12}$.
The latter was discovered shortly before in experiments at the synchrocyclotron in Dubna under the direction of M.G.~Meshcheryakov~\cite{Mesheryakov, Neganov} in the form of an intense peak in the cross-section of the reaction
\begin{equation}\label{eq1}
p + p \to d + \pi^{+},
\end{equation}
Partial wave analysis of reaction~\eqref{eq1} and elastic $NN$ scattering determined the characteristics of the dibaryon states forming the resonance in reaction~\eqref{eq1}~\cite{Arndt1993, Oh}.
The same analysis showed that the observed dibaryon states had the character of true resonances, exhibiting, in particular, adequate behavior on the Argand plane.

The classification~\cite{Dyson} predicted the resonant state of $D_{03}$ with a mass of about 2350~MeV.
The long-term search for this resonance, described in Clement's review~\cite{Clement}, led to its discovery at the WASA-at-COSY setup~\cite{Adlarson2011} in reasonable agreement with the prediction~\cite{Dyson}.
The resonance was investigated not only in the process of inelastic $np$ collision,
\begin{equation}\label{eq2}
n + p \to D_{03} \to d + \pi^{0} \pi^{0},
\end{equation}
but also in quasi-free $np$ scattering~\cite{Adlarson2014}.
Measurement of the $np$ analyzing power showed the presence of a pole in $^3\!D_3$--$^3\!G_3$ waves, confirming the true resonant character of the $D_{03}$ dibaryon.

The isotensor dibaryon $D_{21}$ was discovered relatively recently~\cite{Adlarson2018}, also at WASA-at-COSY.

The search for the not yet observed $D_{30}$ dibaryon, the last of those predicted in~\cite{Dyson} for states with zero strangeness, showed in the experiment~\cite{Adlarson2016} that the upper limit of its excitation probability is three to four orders of magnitude lower than the $D_{03}$ excitation probability.
Such a low cross-section of the corresponding reaction is due to the need to ensure the preservation of isospin in the $pp \to D_{30}\pi^{0}\pi^{0} \to pp$ process by an additional generation of pions.
The small cross-section of such a generation leads to a small probability of $D_{30}$ observation.

Thus, modern data generally confirm the SU(6) systematization of the dibaryons with zero strangeness~\cite{Dyson}.
However, the limitations of this systematization should not be neglected.
Its development requires obtaining new data, in particular, the study of dibaryon resonances with the formation of an $S$-wave diproton ${pp}_s$:
\begin{equation}\label{eq3}
p + p \to D_{IJ} \to \{p p\}_s + \pi^{0},
\end{equation}
Investigation~\cite{Komarov2016} of this process at the ANKE-COSY facility revealed existence of two resonant states $D_{10}$ and $D_{12}$ with a mass of $2.2$~GeV and a decay mode of $\{p p\}_s \pi^{0}$.

Theoretical analysis of the dibaryon resonance phenomenon in the meson-baryon concept has shown that in this approach, dibaryon resonances are systems that are obviously more complex than pairs of non-interacting baryons.
The necessity of taking into account the interaction of baryons in the intermediate state was proved quantitatively in the works of Gal and Garcilazo~\cite{Gal}.

The capabilities of modern QCD-based quark models in quantitative interpretation of dibaryon resonances were demonstrated in~\cite{Dong}, where not only the relatively small full width of the $D_{03}(2380)$ state was reproduced, but also the experimentally known widths of several modes of its decay.
Calculations were carried out in the model of coupled channels $\Delta\Delta + CC$, where the $\Delta\Delta$ channel corresponds to the $t$-channel excitation of two $\Delta(1232)$ baryons, and the $CC$ channel is a six-quark configuration with a hidden color.
The results indicate the dominant contribution of the $CC$ component, which leads the authors to conclude that the $D_{03}(2380)$ dibaryon resonance is an exotic state with six-quark dominance.

The study of the mechanism for excitation of dibaryon resonance is also of fundamental importance for understanding their physical nature.
Historically, such a mechanism, in one form or another, was a nucleon-nucleon collision with a total energy of $\sqrt{s}$ equal to the mass of the excited resonance.
However, another mechanism is also possible, which is coherent excitation of the deuteron to resonant state.
This possibility was shown in~\cite{Komarov2018}, where a deuteron was excited by a meson exchange with a fast proton inelastically scattered at small angles:
\begin{equation}\label{eq4}
p + d \to p + D_{03} \to p + d +\pi^0 \pi^0 ,
\end{equation}
Theoretical analysis of this experiment was done in~\cite{Tursunbaev} in the frame of a theoretical model involving $D(2380)$ excitation in the intermediate state.

Fundamental importance of the dibaryon resonance phenomenon for development of nuclear physics appeared especially clear in the last two decades with the development of a new QCD-motivated approach to the $NN$ intermediate energy scattering, a dibaryon model~\cite{Kukulin}.
This novel model created by the physicists of Moscow State University replaces the $t$-channel meson exchanges in the traditional $NN$-potential models by the $s$-channel mechanism of the dibaryon exchange between the overlapping nucleons together with the peripheral one-pion $t$-channel exchange at long distances.
It was shown in a large set of calculations that the dibaryon degrees of freedom are appropriate to effectively take into account the inner structure of the nucleons in the $NN$ scattering processes.

The above shows that the study of dibaryon resonances, which arose about 70 years ago, presents a wide range of experimental and theoretical problems.
At the same time, the first stage of the study is always the discovery of the very fact of the resonant behavior of the differential sections of the inelastic $NN$ interaction.
This behavior of reaction~\eqref{eq1} has been studied in detail in numerous experiments at energies $\sqrt{s} = 2.1$--$2.4$~GeV, and the energy dependence of the differential cross-section at small angles in the c.m.s.\ has been measured up to 5.0~GeV~\cite{Anderson}.

The reaction
\begin{equation}\label{eq5}
p + p \to \{p p\}_s + \pi^0
\end{equation}
was studied at the ANKE-COSY setup in the energy range $\sqrt{s} = 2.1$--$2.4$~GeV.
The manifestation of heavier baryon resonaces $N^*$ in the dibaryon structure can be observed with increase of  the dibaryon resonance mass,
so it seems appropriate to advance to higher energies in this task.
The aim of this work was to search for study the resonant behavior of the isotriplet interaction of nucleons~\eqref{eq5} leading to the formation of a single pion and an $S$-wave diproton in the range of $\sqrt{s} = 2.48$--$2.91$~GeV.
Experimental equipment and measurement procedure are described in the next section.
Further sections provide an analysis of the data, obtained results and discussion.
The final section summarizes the above.

\section{Measurements}

The reaction $pp \to \{p p\}_s \pi^0$ has been measured with the ANKE-COSY spectrometer~\cite{Barsov} at 5 proton beam kinetic energies in the range $T_p = 1.6-2.4$~GeV.
These new data appreciably add to the previously published results~\cite{Dymov2006, Kurbatov2008, Komarov2016}.

\begin{figure*}[htbp]\centering
\includegraphics[width=0.8\textwidth]{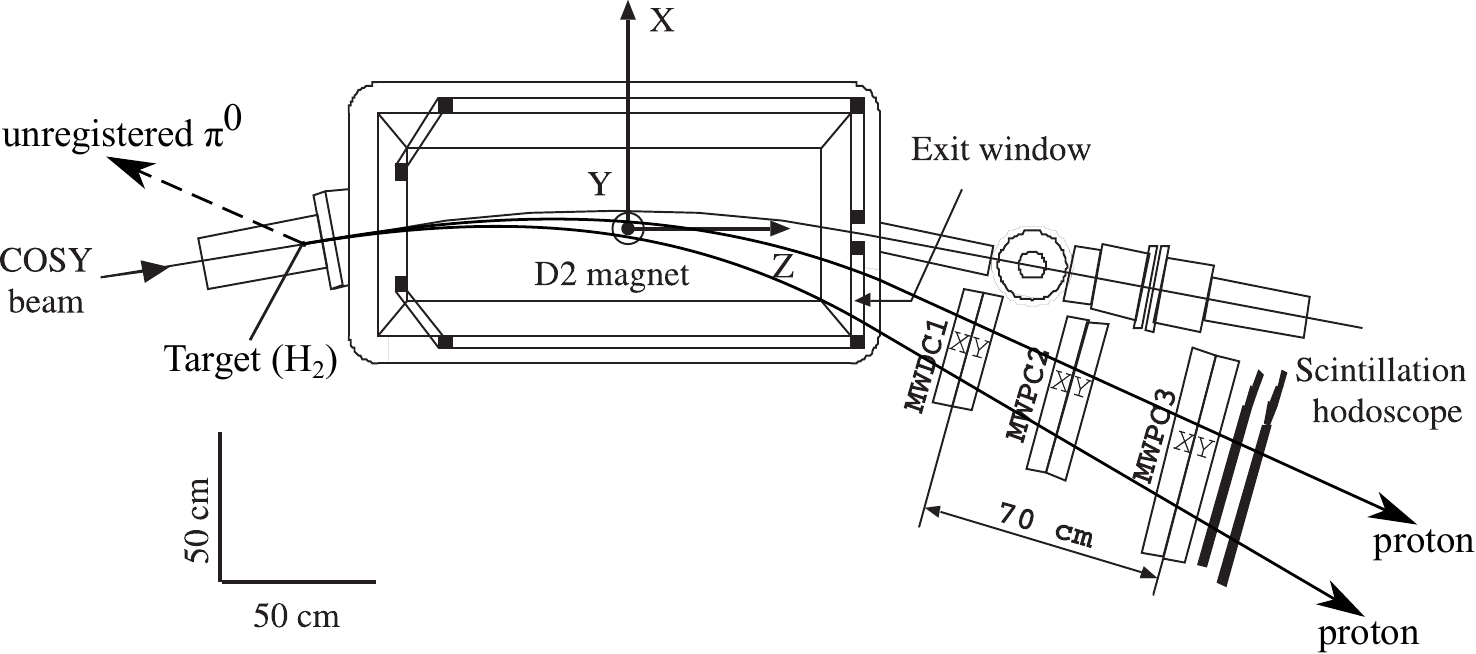}
\caption{Scheme of the ANKE spectrometer.}
\label{ANKE}
\end{figure*}

Figure~\ref{ANKE} shows a scheme of the spectrometer: the beam pipe, the main spectrometric magnet D2 and the ANKE forward detector that was used in the experiment.
The proton beam interacts with a hydrogen cluster-jet target, the secondary protons produced by the interaction are deflected by the spectrometric magnet and detected by the multi-wire chambers and scintillation counters of the ANKE forward detector.
Track coordinates are measured by a set of multi-wire chambers, while scintillation counters measure ionization losses and particle times of flight.
These data make it possible to measure particle trajectories and momenta.
The experimental setup is described in more detail in our earlier papers~\cite{Dymov2006, Tsirkov}.

Data acquisition was activated by the registration of at least one particle with a momentum above $0.6$~GeV/c.
Various triggers were used in the experiment to pre-select and record single-track and double-track events.

\section{Analysis}

The first step in identifying the $pp \to \{p p\}_s \pi^0$ reaction was to select two coincident protons from all detected pairs of positively charged particles.
The time-of-flight method for identifying proton pairs is well suited for this purpose.
A more detailed description of the procedure is given in~\cite{Tsirkov}.

\begin{figure}[htbp]\centering
\includegraphics[width=1\columnwidth]{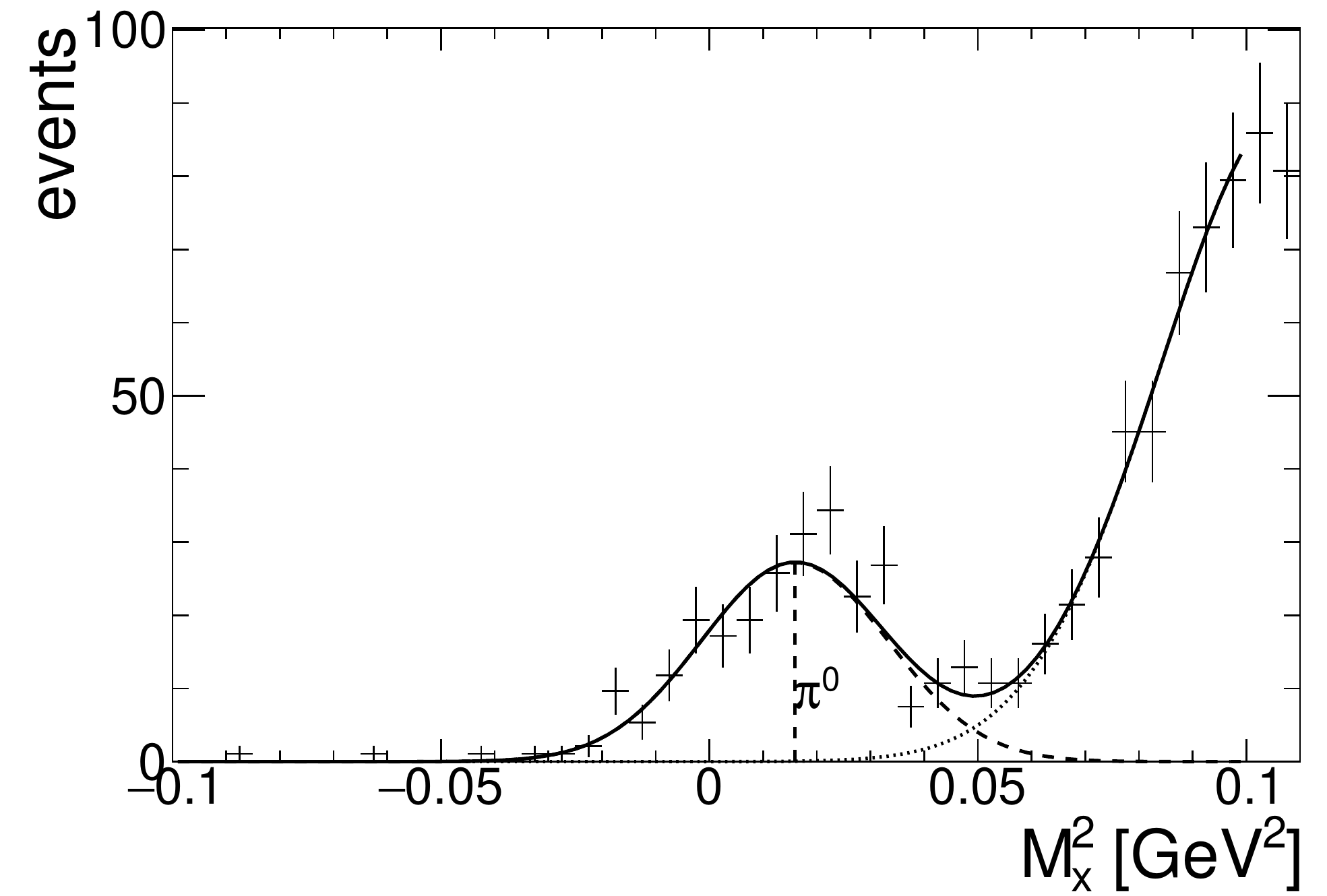}
\caption{Distribution of missing mass squared, an example for 1.6~GeV.}
\label{missing_mass}
\end{figure}

After selecting candidates for the $pp \to pp X$ reaction, we could use the information about the momenta of both final protons to reconstruct the complete kinematics of the process for each event.
In order to select the $^1\!S_0$ state of the pairs, only proton pairs with an excitation energy $E_{pp} < 3$~MeV were chosen.
After that, the candidates for further processing were selected based on the missing mass criterion (see Fig.~\ref{missing_mass} as an example).
Events near the single-pion peak position were accepted for further processing after subtracting a small contribution from the multi-meson production background.
The background contribution was evaluated by fitting various shapes to the data: Gaussian or exponential double-meson tail with or without constant random-coincidence background.
The resulting systematic error amounted to 3\%.
The kinematic fitting technique was applied to the events within the single-pion peak to improve the momentum precision.

\begin{figure}[htbp]\centering
\includegraphics[width=1\columnwidth]{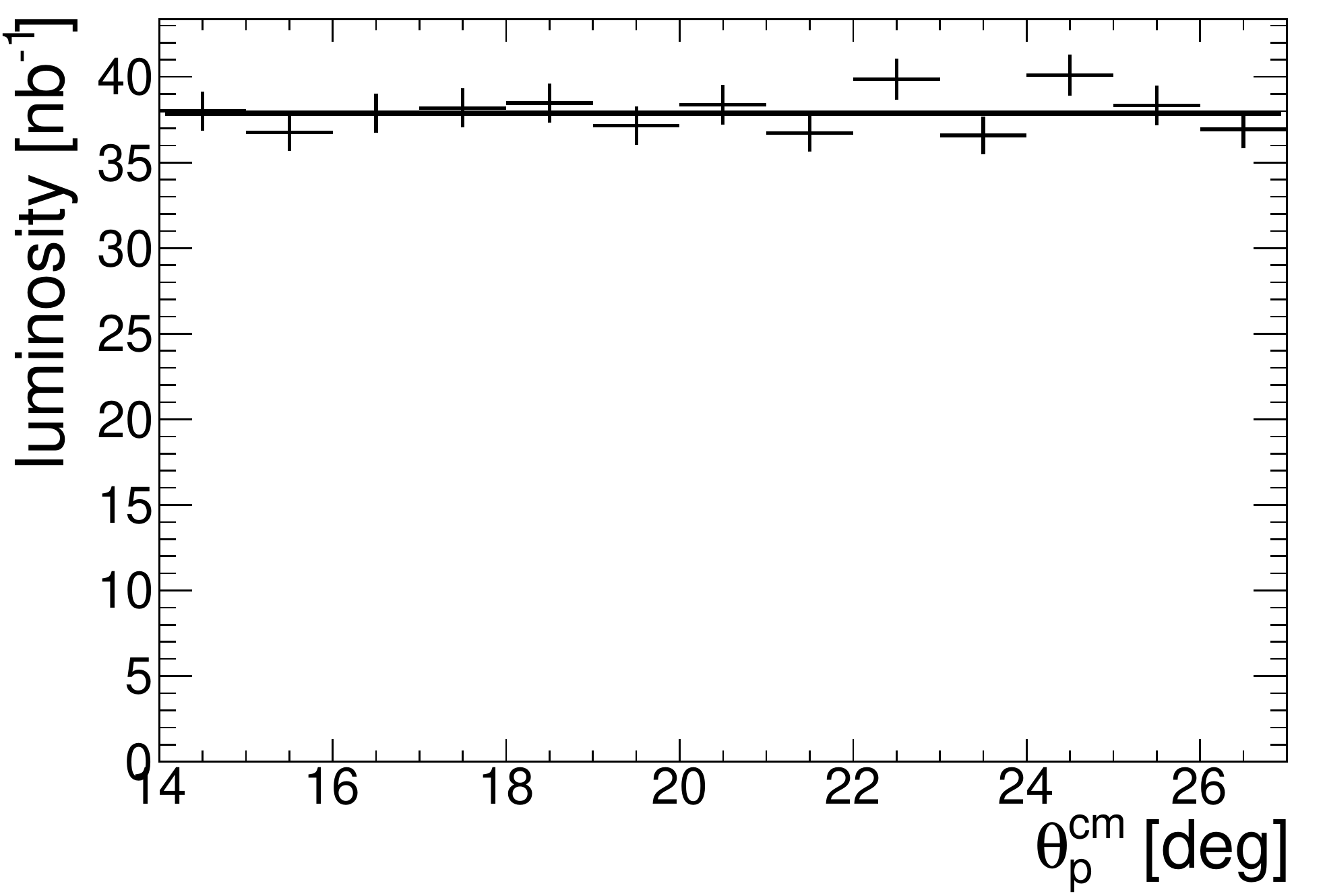}
\caption{Luminosity estimations for various angular intervals fitted by a constant, an example for 1.6~GeV.}
\label{luminosity}
\end{figure}

The luminosity was measured by parallel registration of the single track events from the proton-proton elastic scattering.
For that, the forward detector angular acceptance for the $pp \to pp$ reaction was split into $1^{\circ}$ intervals for $\theta_p^\mathrm{cm}$, and the luminosity was estimated for each interval using $pp \to pp$ cross-sections from the SAID solution SM16~\cite{Workman}, then the obtained anguar distribution was fitted by a constant.
An example of such a fit is presented in Fig.~\ref{luminosity}.
The resulting integral luminosities varied from $1.6$~nb$^{-1}$ to $36$~nb$^{-1}$ for various energies.
The statistical error was negligible due to a large number of $pp$-elastic events, the SAID systematics was considered to be 4\%, the systematics associated with the acceptance uncerainties at the detector edges and registration inefficiencies amounted to 1--3\% at various energies.

\section{Results}

\begin{figure}[htbp]\centering
\includegraphics[width=1\columnwidth]{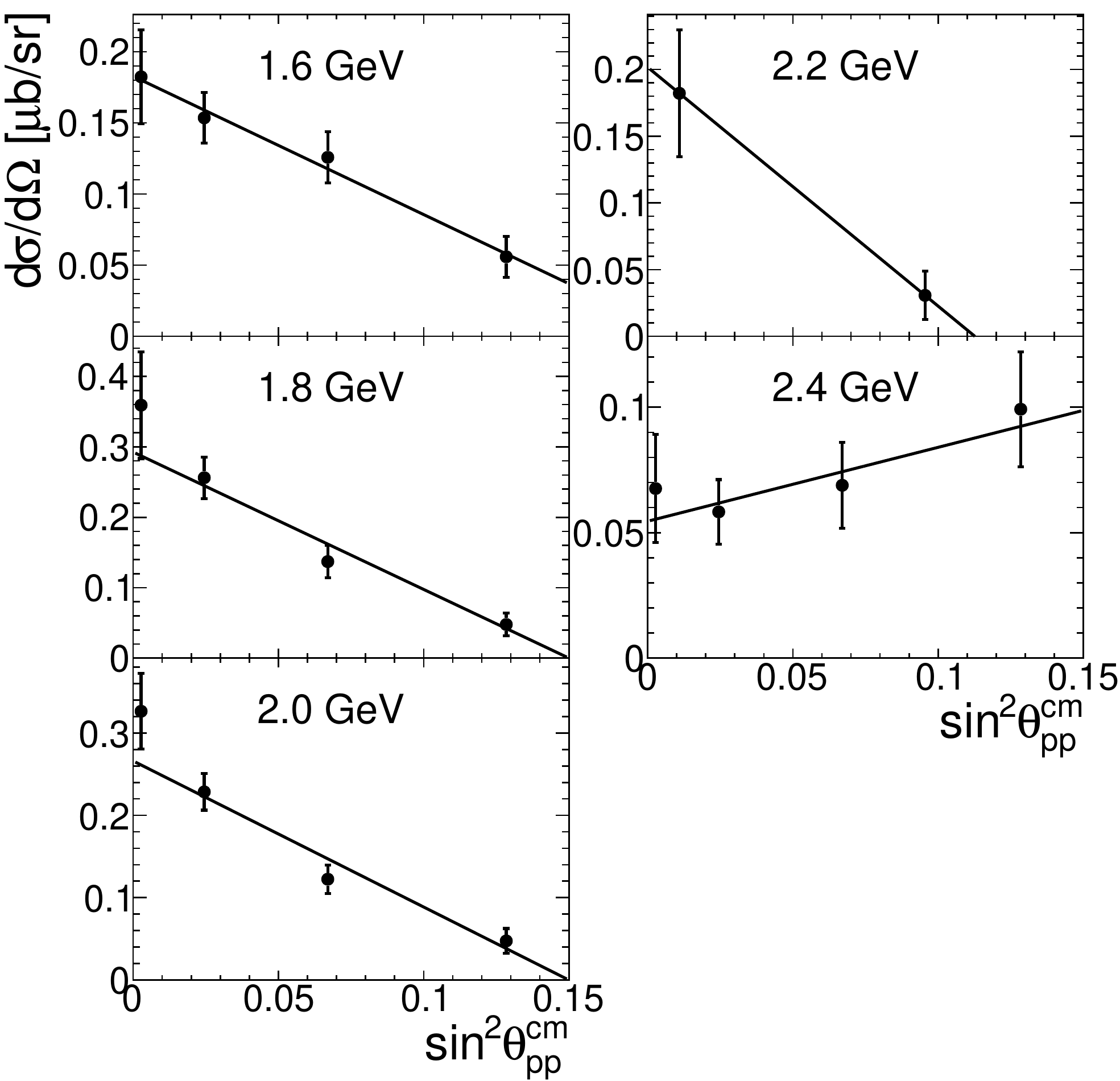}
\caption{Angular dependence of the differential reaction cross-section for the $pp \to \{p p\}_s \pi^0$ reaction}
\label{data_2013}
\end{figure}

\begin{figure}[htbp]\centering
\includegraphics[width=1\columnwidth]{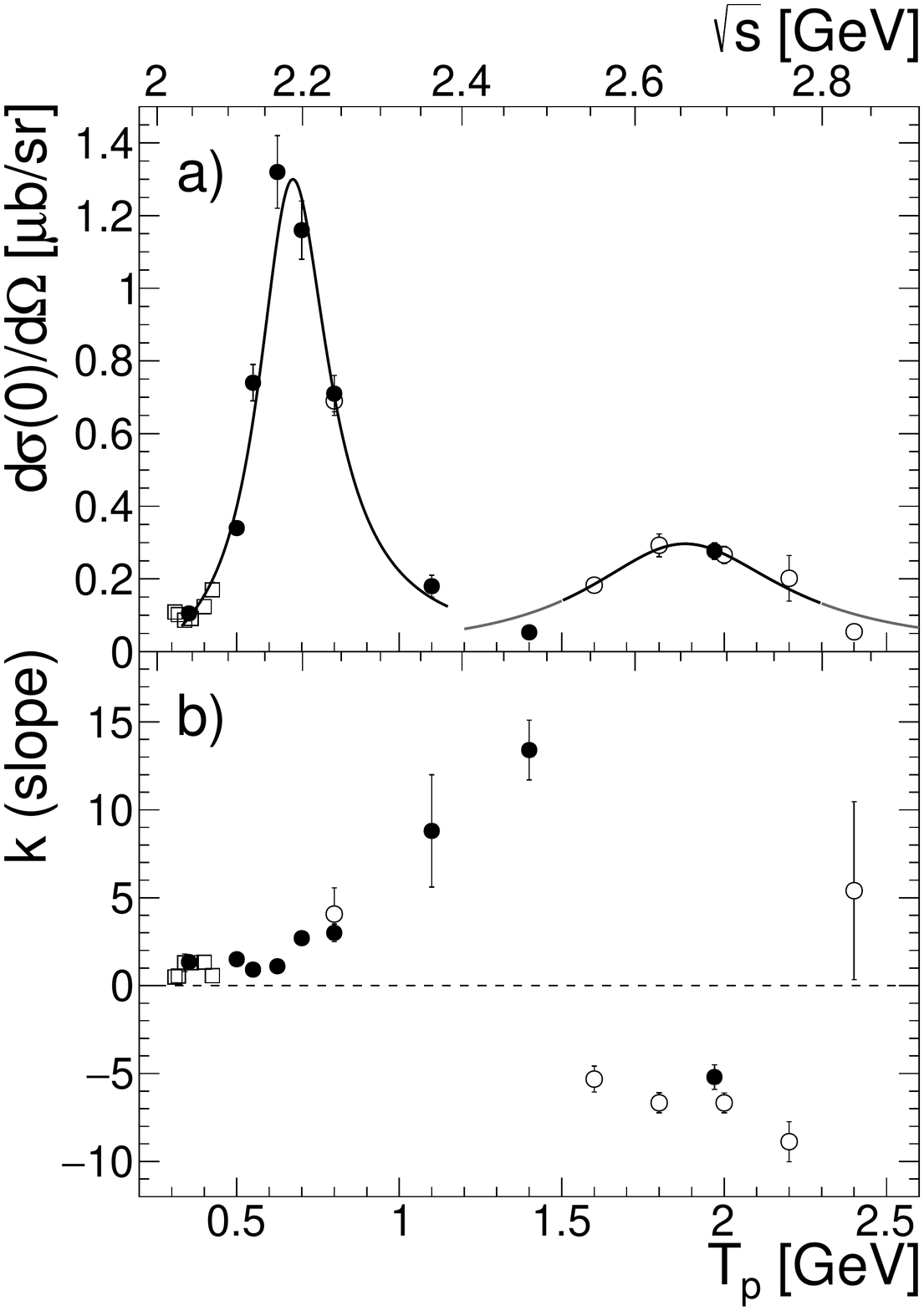}
\caption{
    Energy dependence of (a) the differential cross-section at the zero angle $d\sigma(0)/d\Omega$ and (b) the slope of the differential cross-section $k$ for the $pp \to \{p p\}_s \pi^0$ reaction.
    Open squares are experimental WASA values from~\cite{Bilger}, full circles are ANKE values from~\cite{Dymov2006, Kurbatov2008, Komarov2016}, and open circles are the current ANKE data.
}
\label{final_energy_dependence}
\end{figure}

To estimate the angular dependence of the $pp \to \{p p\}_s \pi^0$ differential cross-section, the events were divided into four intervals by $\theta_{pp}$: 0$^{\circ}$--5$^{\circ}$, 5$^{\circ}$--10$^{\circ}$, 10$^{\circ}$--15$^{\circ}5$, and 15$^{\circ}$--20$^{\circ}$ (Fig.~\ref{data_2013}) and fitted with the function
\begin{equation}\label{eq6}
 \frac{d\sigma}{d\Omega} = \frac{d\sigma(0)}{d\Omega} (1 + k\sin^{2}\theta^\mathrm{cm}_{pp}),
\end{equation}
where $d\sigma(0)/d\Omega$ is the differential cross-section at the zero angle (also called forward cross-section), $k$ is the slope parameter.
The fit results are presented in Fig.~\ref{final_energy_dependence} and Table~\ref{tab_results}.

\newcommand{\ps}{\phantom{$-$}}
\newcommand{\pd}{\phantom{$0$}}
\begin{table*}[htbp]\centering
\caption{Values of the differential cross-section at the zero angle $d\sigma(0)/d\Omega$ and the slope of the differential cross-section $k$ measured at various energies.}
\label{tab_results}
\begin{ruledtabular}
\begin{tabular}{ccccc}
Beam time & $T_p$ [GeV] & $\sqrt{s}$ [GeV] & $d\sigma(0)/d\Omega$ [$\mu$b/sr] & $k$ \\ \hline
2008 data & 1.4         & 2.48             & 0.053 $\pm$ 0.004 & \ps13.4 $\pm$ 1.7\pd \\
          & \pd1.97     & 2.69             & 0.277 $\pm$ 0.023 & $-$5.2 $\pm$ 0.7 \\ \hline
          & 1.6         & 2.55             & 0.183 $\pm$ 0.019 & $-$5.3 $\pm$ 0.7 \\
          & 1.8         & 2.63             & 0.293 $\pm$ 0.031 & $-$6.7 $\pm$ 0.6 \\
2013 data & 2.0         & 2.70             & 0.266 $\pm$ 0.023 & $-$6.7 $\pm$ 0.6 \\
          & 2.2         & 2.77             & 0.202 $\pm$ 0.063 & $-$8.9 $\pm$ 1.1 \\
          & 2.4         & 2.83             & 0.055 $\pm$ 0.014 &   \ps5 $\pm$ 5   \\
\end{tabular}
\end{ruledtabular}
\end{table*}

The resulting values of $d\sigma(0)/d\Omega$ presented in Fig.~\ref{final_energy_dependence} (a) form a peak around $T_p = 1.9$~GeV.
Thus, we made a combined fit of the forward cross-sections published in~\cite{Kurbatov2008} and the current data in the central part of the peak ($T_p = 1.6$--$2.2$~GeV) to exclude the extreme points where some contribution of interference with nearby peaks is possible.
The data were fitted by the Breit-Wigner function
\begin{equation}\label{eq7}
\frac{d\sigma(0)}{d\Omega} = \frac{N}{(\sqrt{s} - E_0)^2 + \Gamma^2/4}.
\end{equation}
where $N$ is the peak magnitude, $E_0$ is the peak position and $\Gamma$ is the peak width.
This results in the magnitude $N$ of $0.30 \pm 0.03$~$\mu$b/sr, the mass $E_0$ of $2.654 \pm 0.013$~GeV, and the width $\Gamma$ of $0.26 \pm 0.07$~GeV.
The observed new resonance may be denoted as $D(2650)$.

In Fig.~\ref{final_energy_dependence} (b) the energy dependence of the slope of the differential cross-section for the $pp \to \{p p\}_s \pi^0$ reaction is shown.
Note that the slope changes its sign in the region of the observed peak.

\section{Discussion}

In the meson-baryon approach the dibaryon resonance can be considered as a pair of interacting baryons of two types $NB^*$ or $B^{*}_{1} B^{*}_{2}$, where $N$ is the nucleon and $B^*$, $B^{*}_{1}$, $B^{*}_{2}$ are different excited baryons.
Masses $M_R$ of the quasi-bound resonant states are close to a sum of the baryon masses and differ from them due to the attractive interaction between the baryons and the kinetic energy of their relative motion.
These quantities depend on the particular intermediate state and have values of the order of tens of MeV.

\begin{figure}[htbp]\centering
\includegraphics[width=0.4\textwidth]{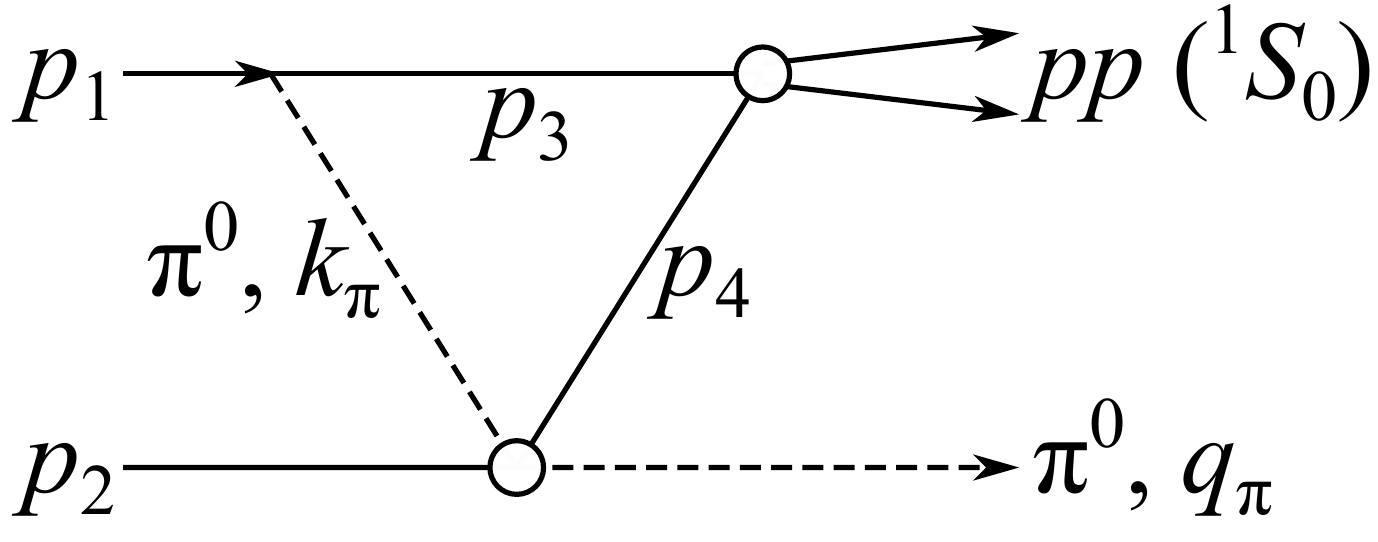}
\caption{The OPE mechanism of the reaction $pp \to \{p p\}_s \pi^0$.}
\label{OPE_mechanism}
\end{figure}

In~\cite{Komarov2016} the dibaryon resonance at 2.2~GeV was interpreted as a resonance in the $P$-wave state of the $\Delta(1232)N$ pair.
It is reasonable to suppose that the new $D(2650)$ resonance is a resonant state of a pair consisting of $N$ and one of the heavier excited $\Delta$ states: $\Delta(1620)$ or $\Delta(1700)$.
To check this assumption, the one-pion exchange (OPE) model can be used (see Fig.~\ref{OPE_mechanism}).
This model was suggested and succesfully applied for the reaction $pp\to d\pi^+$ by T.~Yao~\cite{Yao}.
A proper modification of this model for the reaction $pp \to \{p p\}_s \pi^0$ was done in~\cite{Uzikov2019,*Uzikov2008}.
This triangle diagram of the one-pion exchange does not involve any excited baryon explicitly but supposes their contribution through the corresponding resonance in the pion-nucleon elastic scattering.

\begin{figure}[htbp]\centering
\includegraphics[width=1\columnwidth]{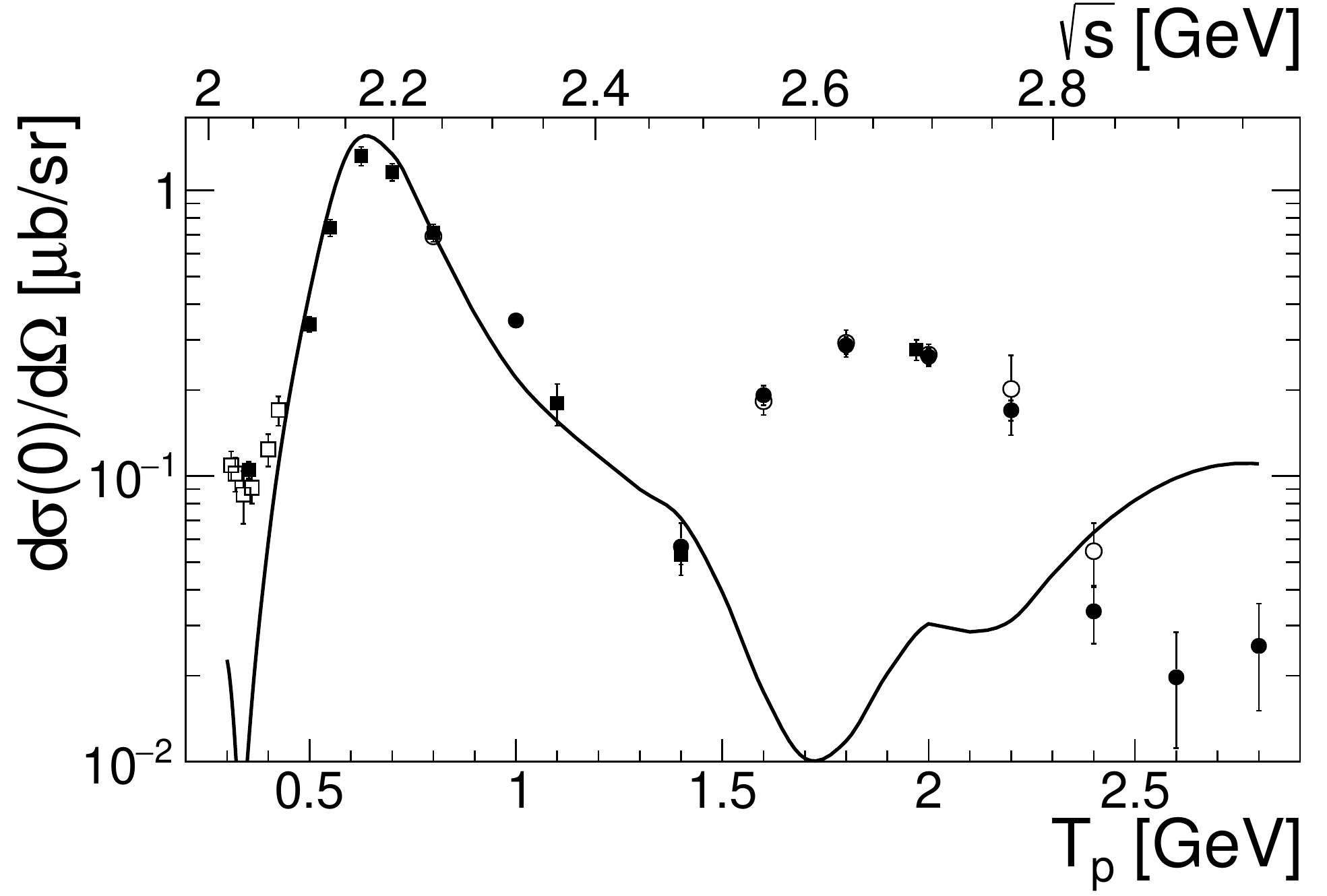}
\caption{
    The result of calculations of the $pp \to \{p p\}_s \pi^0$ differential cross-section at $\theta = 0$ within the OPE model~\cite{Uzikov2019} multiplied by the normalization factor 0.4 (full line).
    The legend for the data points is the same as in Fig.~\ref{final_energy_dependence}.
}
\label{OPE_model}
\end{figure}

Figure~\ref{OPE_model} shows that the OPE model successfully describes the shape of the energy dependence of the differential cross section at the zero angle in the region of the peak observed at $T_p = 0.650$~GeV corresponding to the $\Delta(1232)$-isobar excitation in this reaction.
However, it completely disagrees with the data in the region of the new peak at the energy of excitation of the higher $\Delta$ and $N^{*}$ baryons.
It allows one to conclude that the contribution of the $NB^{*}$-like structure in the observed peak is negligible.

The dibaryon structure with two excited baryons at the lowest excitations (higher than $\Delta(1232) \Delta(1232)$ assumed for $D(2380)$) is $\Delta(1232) N^{*}(1440)$ pair.
According to estimation of the resonance mass as a sum of the masses of these excited baryons, it is $M_R \approx 2.67$~GeV suitable for our resonance.
Taking into account conservation of the total angular momentum, $P$–parity and also requirement of the Pauli principle for the $pp$ state, one should assume a $P$-odd state of the internal motion in this $\Delta(1232) N^{*}(1440)$ pair.
The isospin of this state is $T=1$, as it is fixed by the isospin of the $pp$-state and isospin conservation in this reaction.
Here, for the $D(2650)$ state, only even values of the full angular momentum $J = 0$, $2$, $4$, etc.\ are allowed.
The observed change of sign of the slope parameter in the angular distribution definitely shows the cardinal difference in the structure of the resonances at 2.20~GeV and 2.65~GeV.
Specific behavior of the slope parameter at 2.20~GeV dibaryon resonance was explained in~\cite{Uzikov2017} by the interference between two states, $^3\!P_2d$ and $^3\!P_0s$.
The observed behavior of the slope parameter in the $D(2650)$ testifies the absence of a similar effect.
It is worth noting that the assumed $\Delta(1232)N^{*}(1440)$ structure of the observed resonance indicates a possibility of its significant decay mode with the production of two and even three pions.
It is natural since the $\Delta(1232)$-isobar has a more than 99\% branching ratio for the $N\pi$ decay mode and the Roper baryon $N^{*}(1440)$ has a 60--70\% branching ratio for the $N\pi$ decay mode and 30--40\% for the $N\pi\pi$ mode~\cite{Zyla}.

\section{Summary}

The measured differential cross section of the single-pion production in the proton-proton collisions accompanied by the $^1\!S_0$ diproton forward emission reveals a clear peak in the proton kinetic energy 1.9~GeV.
This peak may be attributed to excitation of the dibaryon resonance $D(2650)$ with the mass of $2.652 \pm 0.005$~GeV and width $\Gamma$ of $0.26 \pm 0.02$~GeV.
The angular dependence of the differential cross-section has shown a change in the sign of the slope parameter between the $D(2220)$ and $D(2650)$ dibaryon resonances, so that in the $D(2650)$ energy region the cross-section has a maximum instead of a minimum at the zero angle.
The calculations in the frame of the one-pion-exchange model shows that the observed cross-section behavior is inconsistent with the $NB^{*}$-type structure.
The resonance may be considered to have the $\Delta(1232)N^{*}(1440)$ structure with the $P$-odd state of the internal movement.
The following study of the resonance requires primarily the measurement of the full angular distribution of the differential cross section, the analyzing powers and the branching ratios for its two- and three-pion decay modes.

\begin{acknowledgments}
The experiment was performed with the ANKE spectrometer in the COSY storage ring at Forschungszentrum Julich (Germany). We acknowledge the contributions by the COSY accelerator crew and other members of the ANKE collaboration.
\end{acknowledgments}

\bibliographystyle{apsrev4-2}
\bibliography{pp_pppi0}

\begin{thebibliography}{28}%
\makeatletter
\providecommand \@ifxundefined [1]{%
 \@ifx{#1\undefined}
}%
\providecommand \@ifnum [1]{%
 \ifnum #1\expandafter \@firstoftwo
 \else \expandafter \@secondoftwo
 \fi
}%
\providecommand \@ifx [1]{%
 \ifx #1\expandafter \@firstoftwo
 \else \expandafter \@secondoftwo
 \fi
}%
\providecommand \natexlab [1]{#1}%
\providecommand \enquote  [1]{``#1''}%
\providecommand \bibnamefont  [1]{#1}%
\providecommand \bibfnamefont [1]{#1}%
\providecommand \citenamefont [1]{#1}%
\providecommand \href@noop [0]{\@secondoftwo}%
\providecommand \href [0]{\begingroup \@sanitize@url \@href}%
\providecommand \@href[1]{\@@startlink{#1}\@@href}%
\providecommand \@@href[1]{\endgroup#1\@@endlink}%
\providecommand \@sanitize@url [0]{\catcode `\\12\catcode `\$12\catcode
  `\&12\catcode `\#12\catcode `\^12\catcode `\_12\catcode `\%12\relax}%
\providecommand \@@startlink[1]{}%
\providecommand \@@endlink[0]{}%
\providecommand \url  [0]{\begingroup\@sanitize@url \@url }%
\providecommand \@url [1]{\endgroup\@href {#1}{\urlprefix }}%
\providecommand \urlprefix  [0]{URL }%
\providecommand \Eprint [0]{\href }%
\providecommand \doibase [0]{https://doi.org/}%
\providecommand \selectlanguage [0]{\@gobble}%
\providecommand \bibinfo  [0]{\@secondoftwo}%
\providecommand \bibfield  [0]{\@secondoftwo}%
\providecommand \translation [1]{[#1]}%
\providecommand \BibitemOpen [0]{}%
\providecommand \bibitemStop [0]{}%
\providecommand \bibitemNoStop [0]{.\EOS\space}%
\providecommand \EOS [0]{\spacefactor3000\relax}%
\providecommand \BibitemShut  [1]{\csname bibitem#1\endcsname}%
\let\auto@bib@innerbib\@empty
\bibitem [{\citenamefont {Gal}\ and\ \citenamefont {Garcilazo}(2014)}]{Gal}%
  \BibitemOpen
  \bibfield  {author} {\bibinfo {author} {\bibfnamefont {A.}~\bibnamefont
  {Gal}}\ and\ \bibinfo {author} {\bibfnamefont {H.}~\bibnamefont
  {Garcilazo}},\ }\href {https://doi.org/10.1016/j.nuclphysa.2014.02.019}
  {\bibfield  {journal} {\bibinfo  {journal} {Nucl. Phys. A}\ }\textbf
  {\bibinfo {volume} {928}},\ \bibinfo {pages} {73} (\bibinfo {year}
  {2014})}\BibitemShut {NoStop}%
\bibitem [{\citenamefont {Dong}\ \emph {et~al.}(2016)\citenamefont {Dong},
  \citenamefont {Huang}, \citenamefont {Shen},\ and\ \citenamefont
  {Zhang}}]{Dong}%
  \BibitemOpen
  \bibfield  {author} {\bibinfo {author} {\bibfnamefont {Y.}~\bibnamefont
  {Dong}}, \bibinfo {author} {\bibfnamefont {F.}~\bibnamefont {Huang}},
  \bibinfo {author} {\bibfnamefont {P.}~\bibnamefont {Shen}},\ and\ \bibinfo
  {author} {\bibfnamefont {Z.}~\bibnamefont {Zhang}},\ }\href
  {https://doi.org/10.1103/PhysRevC.94.014003} {\bibfield  {journal} {\bibinfo
  {journal} {Phys. Rev. C}\ }\textbf {\bibinfo {volume} {94}},\ \bibinfo
  {pages} {014003} (\bibinfo {year} {2016})}\BibitemShut {NoStop}%
\bibitem [{\citenamefont {Dyson}\ and\ \citenamefont {Xuong}(1964)}]{Dyson}%
  \BibitemOpen
  \bibfield  {author} {\bibinfo {author} {\bibfnamefont {F.~J.}\ \bibnamefont
  {Dyson}}\ and\ \bibinfo {author} {\bibfnamefont {N.-H.}\ \bibnamefont
  {Xuong}},\ }\href {https://doi.org/10.1103/PhysRevLett.13.815} {\bibfield
  {journal} {\bibinfo  {journal} {Phys. Rev. Lett.}\ }\textbf {\bibinfo
  {volume} {13}},\ \bibinfo {pages} {815} (\bibinfo {year} {1964})}\BibitemShut
  {NoStop}%
\bibitem [{\citenamefont {Mescheryakov}\ and\ \citenamefont
  {Neganov}(1955)}]{Mesheryakov}%
  \BibitemOpen
  \bibfield  {author} {\bibinfo {author} {\bibfnamefont {M.~G.}\ \bibnamefont
  {Mescheryakov}}\ and\ \bibinfo {author} {\bibfnamefont {B.~S.}\ \bibnamefont
  {Neganov}},\ }\href@noop {} {\bibfield  {journal} {\bibinfo  {journal} {Sov.
  Phys. Dokl.}\ }\textbf {\bibinfo {volume} {100}},\ \bibinfo {pages} {677}
  (\bibinfo {year} {1955})}\BibitemShut {NoStop}%
\bibitem [{\citenamefont {Neganov}\ and\ \citenamefont
  {Parfenov}(1958)}]{Neganov}%
  \BibitemOpen
  \bibfield  {author} {\bibinfo {author} {\bibfnamefont {B.~S.}\ \bibnamefont
  {Neganov}}\ and\ \bibinfo {author} {\bibfnamefont {L.~B.}\ \bibnamefont
  {Parfenov}},\ }\href@noop {} {\bibfield  {journal} {\bibinfo  {journal} {JETP
  Lett.}\ }\textbf {\bibinfo {volume} {7}},\ \bibinfo {pages} {528} (\bibinfo
  {year} {1958})}\BibitemShut {NoStop}%
\bibitem [{\citenamefont {Arndt}\ \emph {et~al.}(1993)\citenamefont {Arndt},
  \citenamefont {Strakovsky}, \citenamefont {Workman},\ and\ \citenamefont
  {Bugg}}]{Arndt1993}%
  \BibitemOpen
  \bibfield  {author} {\bibinfo {author} {\bibfnamefont {R.~A.}\ \bibnamefont
  {Arndt}}, \bibinfo {author} {\bibfnamefont {I.~I.}\ \bibnamefont
  {Strakovsky}}, \bibinfo {author} {\bibfnamefont {R.~L.}\ \bibnamefont
  {Workman}},\ and\ \bibinfo {author} {\bibfnamefont {D.~V.}\ \bibnamefont
  {Bugg}},\ }\href {https://doi.org/10.1103/PhysRevC.48.1926} {\bibfield
  {journal} {\bibinfo  {journal} {Phys. Rev. C}\ }\textbf {\bibinfo {volume}
  {48}},\ \bibinfo {pages} {1926} (\bibinfo {year} {1993})}\BibitemShut
  {NoStop}%
\bibitem [{\citenamefont {Oh}\ \emph {et~al.}(1997)\citenamefont {Oh},
  \citenamefont {Arndt}, \citenamefont {Strakovsky},\ and\ \citenamefont
  {Workmanet}}]{Oh}%
  \BibitemOpen
  \bibfield  {author} {\bibinfo {author} {\bibfnamefont {C.~H.}\ \bibnamefont
  {Oh}}, \bibinfo {author} {\bibfnamefont {R.~A.}\ \bibnamefont {Arndt}},
  \bibinfo {author} {\bibfnamefont {I.~I.}\ \bibnamefont {Strakovsky}},\ and\
  \bibinfo {author} {\bibfnamefont {R.~L.}\ \bibnamefont {Workmanet}},\ }\href
  {https://doi.org/10.1103/PhysRevC.56.635} {\bibfield  {journal} {\bibinfo
  {journal} {Phys. Rev. C}\ }\textbf {\bibinfo {volume} {56}},\ \bibinfo
  {pages} {635} (\bibinfo {year} {1997})}\BibitemShut {NoStop}%
\bibitem [{\citenamefont {Clement}(2017)}]{Clement}%
  \BibitemOpen
  \bibfield  {author} {\bibinfo {author} {\bibfnamefont {H.}~\bibnamefont
  {Clement}},\ }\href {https://doi.org/10.1016/j.ppnp.2016.12.004} {\bibfield
  {journal} {\bibinfo  {journal} {Prog. Part. Nucl. Phys.}\ }\textbf {\bibinfo
  {volume} {93}},\ \bibinfo {pages} {196} (\bibinfo {year} {2017})}\BibitemShut
  {NoStop}%
\bibitem [{\citenamefont {Adlarson}\ \emph {et~al.}(2011)\citenamefont
  {Adlarson} \emph {et~al.}}]{Adlarson2011}%
  \BibitemOpen
  \bibfield  {author} {\bibinfo {author} {\bibfnamefont {P.}~\bibnamefont
  {Adlarson}} \emph {et~al.} (\bibinfo {collaboration} {WASA-at-COSY
  Collaboration}),\ }\href {https://doi.org/10.1103/PhysRevLett.106.242302}
  {\bibfield  {journal} {\bibinfo  {journal} {Phys. Rev. Lett.}\ }\textbf
  {\bibinfo {volume} {106}},\ \bibinfo {pages} {242302} (\bibinfo {year}
  {2011})}\BibitemShut {NoStop}%
\bibitem [{\citenamefont {Adlarson}\ \emph {et~al.}(2014)\citenamefont
  {Adlarson} \emph {et~al.}}]{Adlarson2014}%
  \BibitemOpen
  \bibfield  {author} {\bibinfo {author} {\bibfnamefont {P.}~\bibnamefont
  {Adlarson}} \emph {et~al.} (\bibinfo {collaboration} {WASA-at-COSY
  Collaboration and SAID Data Analysis Center}),\ }\href
  {https://doi.org/10.1103/PhysRevLett.112.202301} {\bibfield  {journal}
  {\bibinfo  {journal} {Phys. Rev. Lett.}\ }\textbf {\bibinfo {volume} {112}},\
  \bibinfo {pages} {202301} (\bibinfo {year} {2014})}\BibitemShut {NoStop}%
\bibitem [{\citenamefont {Adlarson}\ \emph {et~al.}(2018)\citenamefont
  {Adlarson} \emph {et~al.}}]{Adlarson2018}%
  \BibitemOpen
  \bibfield  {author} {\bibinfo {author} {\bibfnamefont {P.}~\bibnamefont
  {Adlarson}} \emph {et~al.} (\bibinfo {collaboration} {WASA-at-COSY
  Collaboration and SAID Data Analysis Center}),\ }\href
  {https://doi.org/10.1103/PhysRevLett.121.052001} {\bibfield  {journal}
  {\bibinfo  {journal} {Phys. Rev. Lett.}\ }\textbf {\bibinfo {volume} {121}},\
  \bibinfo {pages} {052001} (\bibinfo {year} {2018})}\BibitemShut {NoStop}%
\bibitem [{\citenamefont {Adlarson}\ \emph {et~al.}(2016)\citenamefont
  {Adlarson} \emph {et~al.}}]{Adlarson2016}%
  \BibitemOpen
  \bibfield  {author} {\bibinfo {author} {\bibfnamefont {P.}~\bibnamefont
  {Adlarson}} \emph {et~al.} (\bibinfo {collaboration} {WASA-at-COSY
  Collaboration}),\ }\href {https://doi.org/10.1016/j.physletb.2016.09.051}
  {\bibfield  {journal} {\bibinfo  {journal} {Phys. Lett. B}\ }\textbf
  {\bibinfo {volume} {762}},\ \bibinfo {pages} {455} (\bibinfo {year}
  {2016})}\BibitemShut {NoStop}%
\bibitem [{\citenamefont {Komarov}\ \emph {et~al.}(2016)\citenamefont
  {Komarov}, \citenamefont {Tsirkov}, \citenamefont {Azaryan}, \citenamefont
  {Bagdasarian}, \citenamefont {Dymov}, \citenamefont {Gebel}, \citenamefont
  {Gou}, \citenamefont {Kacharava}, \citenamefont {Khoukaz}, \citenamefont
  {Kulikov} \emph {et~al.}}]{Komarov2016}%
  \BibitemOpen
  \bibfield  {author} {\bibinfo {author} {\bibfnamefont {V.}~\bibnamefont
  {Komarov}}, \bibinfo {author} {\bibfnamefont {D.}~\bibnamefont {Tsirkov}},
  \bibinfo {author} {\bibfnamefont {T.}~\bibnamefont {Azaryan}}, \bibinfo
  {author} {\bibfnamefont {Z.}~\bibnamefont {Bagdasarian}}, \bibinfo {author}
  {\bibfnamefont {S.}~\bibnamefont {Dymov}}, \bibinfo {author} {\bibfnamefont
  {R.}~\bibnamefont {Gebel}}, \bibinfo {author} {\bibfnamefont
  {B.}~\bibnamefont {Gou}}, \bibinfo {author} {\bibfnamefont {A.}~\bibnamefont
  {Kacharava}}, \bibinfo {author} {\bibfnamefont {A.}~\bibnamefont {Khoukaz}},
  \bibinfo {author} {\bibfnamefont {A.}~\bibnamefont {Kulikov}}, \emph
  {et~al.},\ }\href {https://doi.org/10.1103/PhysRevC.93.065206} {\bibfield
  {journal} {\bibinfo  {journal} {Phys. Rev. C}\ }\textbf {\bibinfo {volume}
  {93}},\ \bibinfo {pages} {065206} (\bibinfo {year} {2016})}\BibitemShut
  {NoStop}%
\bibitem [{\citenamefont {Komarov}\ \emph {et~al.}(2018)\citenamefont
  {Komarov}, \citenamefont {Tsirkov}, \citenamefont {Azaryan}, \citenamefont
  {Bagdasarian}, \citenamefont {Baimurzinova}, \citenamefont {Barsov},
  \citenamefont {Dymov}, \citenamefont {Gebel}, \citenamefont {Hartmann},
  \citenamefont {Kacharava} \emph {et~al.}}]{Komarov2018}%
  \BibitemOpen
  \bibfield  {author} {\bibinfo {author} {\bibfnamefont {V.~I.}\ \bibnamefont
  {Komarov}}, \bibinfo {author} {\bibfnamefont {D.}~\bibnamefont {Tsirkov}},
  \bibinfo {author} {\bibfnamefont {T.}~\bibnamefont {Azaryan}}, \bibinfo
  {author} {\bibfnamefont {Z.}~\bibnamefont {Bagdasarian}}, \bibinfo {author}
  {\bibfnamefont {B.}~\bibnamefont {Baimurzinova}}, \bibinfo {author}
  {\bibfnamefont {S.}~\bibnamefont {Barsov}}, \bibinfo {author} {\bibfnamefont
  {S.}~\bibnamefont {Dymov}}, \bibinfo {author} {\bibfnamefont
  {R.}~\bibnamefont {Gebel}}, \bibinfo {author} {\bibfnamefont
  {M.}~\bibnamefont {Hartmann}}, \bibinfo {author} {\bibfnamefont
  {A.}~\bibnamefont {Kacharava}}, \emph {et~al.},\ }\href
  {https://doi.org/10.1140/epja/i2018-12641-0} {\bibfield  {journal} {\bibinfo
  {journal} {Eur. Phys. J. A}\ }\textbf {\bibinfo {volume} {54}},\ \bibinfo
  {pages} {206} (\bibinfo {year} {2018})}\BibitemShut {NoStop}%
\bibitem [{\citenamefont {Tursunbaev}\ and\ \citenamefont
  {Uzikov}(2020)}]{Tursunbaev}%
  \BibitemOpen
  \bibfield  {author} {\bibinfo {author} {\bibfnamefont {N.}~\bibnamefont
  {Tursunbaev}}\ and\ \bibinfo {author} {\bibfnamefont {Y.}~\bibnamefont
  {Uzikov}},\ }\href {https://doi.org/10.1007/978-3-030-32357-8_76} {\bibfield
  {journal} {\bibinfo  {journal} {Springer Proceedings in Physics}\ }\textbf
  {\bibinfo {volume} {238}} (\bibinfo {year} {2020})}\BibitemShut {NoStop}%
\bibitem [{\citenamefont {Kukulin}\ \emph {et~al.}(2022)\citenamefont
  {Kukulin}, \citenamefont {Pomerantsev}, \citenamefont {Rubtsova},
  \citenamefont {Platonova},\ and\ \citenamefont {Obukhovsky}}]{Kukulin}%
  \BibitemOpen
  \bibfield  {author} {\bibinfo {author} {\bibfnamefont {V.~I.}\ \bibnamefont
  {Kukulin}}, \bibinfo {author} {\bibfnamefont {V.~N.}\ \bibnamefont
  {Pomerantsev}}, \bibinfo {author} {\bibfnamefont {O.~A.}\ \bibnamefont
  {Rubtsova}}, \bibinfo {author} {\bibfnamefont {M.~N.}\ \bibnamefont
  {Platonova}},\ and\ \bibinfo {author} {\bibfnamefont {I.~T.}\ \bibnamefont
  {Obukhovsky}},\ }\Eprint {https://arxiv.org/abs/2205.06377} {arXiv:2205.06377
  [nucl-th]}  (\bibinfo {year} {2022})\BibitemShut {NoStop}%
\bibitem [{\citenamefont {Anderson}\ \emph {et~al.}(1971)\citenamefont
  {Anderson} \emph {et~al.}}]{Anderson}%
  \BibitemOpen
  \bibfield  {author} {\bibinfo {author} {\bibfnamefont {H.~L.}\ \bibnamefont
  {Anderson}} \emph {et~al.},\ }\href {https://doi.org/10.1103/PhysRevD.3.1536}
  {\bibfield  {journal} {\bibinfo  {journal} {Phys. Rev. D}\ }\textbf {\bibinfo
  {volume} {3}},\ \bibinfo {pages} {1536} (\bibinfo {year} {1971})}\BibitemShut
  {NoStop}%
\bibitem [{\citenamefont {Barsov}\ \emph {et~al.}(2001)\citenamefont {Barsov}
  \emph {et~al.}}]{Barsov}%
  \BibitemOpen
  \bibfield  {author} {\bibinfo {author} {\bibfnamefont {S.}~\bibnamefont
  {Barsov}} \emph {et~al.},\ }\href
  {https://doi.org/10.1016/S0168-9002(00)01147-5} {\bibfield  {journal}
  {\bibinfo  {journal} {Nucl. Instrum. Methods Phys. Res. A}\ }\textbf
  {\bibinfo {volume} {462}},\ \bibinfo {pages} {364} (\bibinfo {year}
  {2001})}\BibitemShut {NoStop}%
\bibitem [{\citenamefont {Dymov}\ \emph {et~al.}(2006)\citenamefont {Dymov},
  \citenamefont {B{\"u}scher}, \citenamefont {Gusev}, \citenamefont {Hartmann},
  \citenamefont {Hejny}, \citenamefont {Kacharava}, \citenamefont {Khoukaz},
  \citenamefont {Komarov}, \citenamefont {Kulessa}, \citenamefont {Kulikov}
  \emph {et~al.}}]{Dymov2006}%
  \BibitemOpen
  \bibfield  {author} {\bibinfo {author} {\bibfnamefont {S.}~\bibnamefont
  {Dymov}}, \bibinfo {author} {\bibfnamefont {M.}~\bibnamefont {B{\"u}scher}},
  \bibinfo {author} {\bibfnamefont {D.}~\bibnamefont {Gusev}}, \bibinfo
  {author} {\bibfnamefont {M.}~\bibnamefont {Hartmann}}, \bibinfo {author}
  {\bibfnamefont {V.}~\bibnamefont {Hejny}}, \bibinfo {author} {\bibfnamefont
  {A.}~\bibnamefont {Kacharava}}, \bibinfo {author} {\bibfnamefont
  {A.}~\bibnamefont {Khoukaz}}, \bibinfo {author} {\bibfnamefont
  {V.}~\bibnamefont {Komarov}}, \bibinfo {author} {\bibfnamefont
  {P.}~\bibnamefont {Kulessa}}, \bibinfo {author} {\bibfnamefont
  {A.}~\bibnamefont {Kulikov}}, \emph {et~al.} (\bibinfo {collaboration}
  {ANKE}),\ }\href {https://doi.org/10.1016/j.physletb.2006.03.012} {\bibfield
  {journal} {\bibinfo  {journal} {Phys. Lett. B}\ }\textbf {\bibinfo {volume}
  {635}},\ \bibinfo {pages} {270} (\bibinfo {year} {2006})}\BibitemShut
  {NoStop}%
\bibitem [{\citenamefont {Kurbatov}\ \emph {et~al.}(2008)\citenamefont
  {Kurbatov}, \citenamefont {Büscher}, \citenamefont {Dymov}, \citenamefont
  {Gusev}, \citenamefont {Hartmann}, \citenamefont {Kacharava}, \citenamefont
  {Khoukaz}, \citenamefont {Komarov}, \citenamefont {Kulikov}, \citenamefont
  {Macharashvili} \emph {et~al.}}]{Kurbatov2008}%
  \BibitemOpen
  \bibfield  {author} {\bibinfo {author} {\bibfnamefont {V.}~\bibnamefont
  {Kurbatov}}, \bibinfo {author} {\bibfnamefont {M.}~\bibnamefont {Büscher}},
  \bibinfo {author} {\bibfnamefont {S.}~\bibnamefont {Dymov}}, \bibinfo
  {author} {\bibfnamefont {D.}~\bibnamefont {Gusev}}, \bibinfo {author}
  {\bibfnamefont {M.}~\bibnamefont {Hartmann}}, \bibinfo {author}
  {\bibfnamefont {A.}~\bibnamefont {Kacharava}}, \bibinfo {author}
  {\bibfnamefont {A.}~\bibnamefont {Khoukaz}}, \bibinfo {author} {\bibfnamefont
  {V.}~\bibnamefont {Komarov}}, \bibinfo {author} {\bibfnamefont
  {A.}~\bibnamefont {Kulikov}}, \bibinfo {author} {\bibfnamefont
  {G.}~\bibnamefont {Macharashvili}}, \emph {et~al.},\ }\href
  {https://doi.org/10.1016/j.physletb.2008.01.051} {\bibfield  {journal}
  {\bibinfo  {journal} {Phys. Lett. B}\ }\textbf {\bibinfo {volume} {661}},\
  \bibinfo {pages} {22} (\bibinfo {year} {2008})}\BibitemShut {NoStop}%
\bibitem [{\citenamefont {Tsirkov}\ \emph {et~al.}(2010)\citenamefont
  {Tsirkov}, \citenamefont {Komarov}, \citenamefont {Azaryan}, \citenamefont
  {Chiladze}, \citenamefont {Dymov}, \citenamefont {Dzyuba}, \citenamefont
  {Hartmann}, \citenamefont {Kacharava}, \citenamefont {Khoukaz}, \citenamefont
  {Kulikov} \emph {et~al.}}]{Tsirkov}%
  \BibitemOpen
  \bibfield  {author} {\bibinfo {author} {\bibfnamefont {D.}~\bibnamefont
  {Tsirkov}}, \bibinfo {author} {\bibfnamefont {V.}~\bibnamefont {Komarov}},
  \bibinfo {author} {\bibfnamefont {T.}~\bibnamefont {Azaryan}}, \bibinfo
  {author} {\bibfnamefont {D.}~\bibnamefont {Chiladze}}, \bibinfo {author}
  {\bibfnamefont {S.}~\bibnamefont {Dymov}}, \bibinfo {author} {\bibfnamefont
  {A.}~\bibnamefont {Dzyuba}}, \bibinfo {author} {\bibfnamefont
  {M.}~\bibnamefont {Hartmann}}, \bibinfo {author} {\bibfnamefont
  {A.}~\bibnamefont {Kacharava}}, \bibinfo {author} {\bibfnamefont
  {A.}~\bibnamefont {Khoukaz}}, \bibinfo {author} {\bibfnamefont
  {A.}~\bibnamefont {Kulikov}}, \emph {et~al.},\ }\href
  {https://doi.org/10.1088/0954-3899/37/10/105005} {\bibfield  {journal}
  {\bibinfo  {journal} {J. Phys. G: Nucl. Part. Phys.}\ }\textbf {\bibinfo
  {volume} {37}},\ \bibinfo {pages} {105005} (\bibinfo {year}
  {2010})}\BibitemShut {NoStop}%
\bibitem [{\citenamefont {Workman}\ \emph {et~al.}(2016)\citenamefont
  {Workman}, \citenamefont {Briscoe},\ and\ \citenamefont
  {Strakovsky}}]{Workman}%
  \BibitemOpen
  \bibfield  {author} {\bibinfo {author} {\bibfnamefont {R.~L.}\ \bibnamefont
  {Workman}}, \bibinfo {author} {\bibfnamefont {W.~J.}\ \bibnamefont
  {Briscoe}},\ and\ \bibinfo {author} {\bibfnamefont {I.~I.}\ \bibnamefont
  {Strakovsky}},\ }\href {https://doi.org/10.1103/PhysRevC.94.065203}
  {\bibfield  {journal} {\bibinfo  {journal} {Phys. Rev. C}\ }\textbf {\bibinfo
  {volume} {94}},\ \bibinfo {pages} {065203} (\bibinfo {year}
  {2016})}\BibitemShut {NoStop}%
\bibitem [{\citenamefont {Bilger}\ \emph {et~al.}(2001)\citenamefont {Bilger}
  \emph {et~al.}}]{Bilger}%
  \BibitemOpen
  \bibfield  {author} {\bibinfo {author} {\bibfnamefont {R.}~\bibnamefont
  {Bilger}} \emph {et~al.},\ }\href
  {https://doi.org/10.1016/S0375-9474(01)00800-4} {\bibfield  {journal}
  {\bibinfo  {journal} {Nucl. Phys. A}\ }\textbf {\bibinfo {volume} {693}},\
  \bibinfo {pages} {633} (\bibinfo {year} {2001})}\BibitemShut {NoStop}%
\bibitem [{\citenamefont {Yao}(1964)}]{Yao}%
  \BibitemOpen
  \bibfield  {author} {\bibinfo {author} {\bibfnamefont {T.}~\bibnamefont
  {Yao}},\ }\href {https://doi.org/10.1103/PhysRev.134.B454} {\bibfield
  {journal} {\bibinfo  {journal} {Phys. Rev.}\ }\textbf {\bibinfo {volume}
  {134}},\ \bibinfo {pages} {B454} (\bibinfo {year} {1964})}\BibitemShut
  {NoStop}%
\bibitem [{\citenamefont {Uzikov}(2019)}]{Uzikov2019}%
  \BibitemOpen
  \bibfield  {author} {\bibinfo {author} {\bibfnamefont {Y.}~\bibnamefont
  {Uzikov}},\ }\href {https://doi.org/10.1051/epjconf/201920401015} {\bibfield
  {journal} {\bibinfo  {journal} {EPJ Web of Conferences}\ }\textbf {\bibinfo
  {volume} {204}},\ \bibinfo {pages} {01015} (\bibinfo {year}
  {2019})}\BibitemShut {NoStop}%
\bibitem [{\citenamefont {Uzikov}(2008)}]{Uzikov2008}%
  \BibitemOpen
  \bibfield  {author} {\bibinfo {author} {\bibfnamefont {Y.~N.}\ \bibnamefont
  {Uzikov}},\ }\Eprint {https://arxiv.org/abs/0803.2342} {arXiv:0803.2342
  [nucl-th]}  (\bibinfo {year} {2008})\BibitemShut {NoStop}%
\bibitem [{\citenamefont {Uzikov}(2017)}]{Uzikov2017}%
  \BibitemOpen
  \bibfield  {author} {\bibinfo {author} {\bibfnamefont {Y.}~\bibnamefont
  {Uzikov}},\ }\href {https://doi.org/10.3103/S1062873817060235} {\bibfield
  {journal} {\bibinfo  {journal} {Izv. Ross. Akad. Nauk Ser. Fiz.}\ }\textbf
  {\bibinfo {volume} {81}},\ \bibinfo {pages} {814} (\bibinfo {year}
  {2017})}\BibitemShut {NoStop}%
\bibitem [{\citenamefont {Zyla}\ \emph {et~al.}(2020)\citenamefont {Zyla} \emph
  {et~al.}}]{Zyla}%
  \BibitemOpen
  \bibfield  {author} {\bibinfo {author} {\bibfnamefont {P.~A.}\ \bibnamefont
  {Zyla}} \emph {et~al.} (\bibinfo {collaboration} {Particle Data Group}),\
  }\href {https://doi.org/10.1093/ptep/ptaa104} {\bibfield  {journal} {\bibinfo
   {journal} {Prog. Theor. Exp. Phys.}\ }\textbf {\bibinfo {volume} {2020}},\
  \bibinfo {pages} {083C01} (\bibinfo {year} {2020})}\BibitemShut {NoStop}%
\end{thebibliography}%

\end{document}